\documentclass[manuscript]{acmart}
\usepackage[noend]{algpseudocode}
\usepackage{algorithmicx,algorithm}
\usepackage{url}

\AtBeginDocument{%
  \providecommand\BibTeX{{%
    \normalfont B\kern-0.5em{\scshape i\kern-0.25em b}\kern-0.8em\TeX}}}

\setcopyright{acmcopyright}
\copyrightyear{2021}
\acmYear{2021}
\acmDOI{10.1145/xxxxxxx.xxxxxxx}

\begin{document}

\title{HeadText: Exploring Hands-free Text Entry using Head Gestures by Motion Sensing on a Smart Earpiece}

\author{Songlin Xu}
\affiliation{%
  \institution{University of Science and Technology of China}
  \city{Hefei}
  \country{China}}
\email{xsl314@mail.ustc.edu.cn}

\author{Guanjie Wang}
\affiliation{%
  \institution{University of Science and Technology of China}
  \city{Hefei}
  \country{China}
}

\author{Ziyuan Fang}
\affiliation{%
 \institution{University of Science and Technology of China}
 \city{Hefei}
 \country{China}}

\author{Guangwei Zhang}
\affiliation{%
  \institution{University of Science and Technology of China}
  \city{Hefei}
  \country{China}}

\author{Guangzhu Shang}
\affiliation{%
  \institution{University of Science and Technology of China}
  \city{Hefei}
  \country{China}}

\author{Rongde Lu}
\affiliation{\institution{University of Science and Technology of China}
  \city{Hefei}
  \country{China}}

\author{Liqun He}
\affiliation{\institution{University of Science and Technology of China}
  \city{Hefei}
  \country{China}}
\email{heliqun@ustc.edu.cn}

\renewcommand{\shortauthors}{Xu, et al.}

\begin{abstract}
  We present HeadText, a hands-free technique on a smart earpiece for text entry by motion sensing. Users input text utilizing only 7 head gestures for key selection, word selection, word commitment and word cancelling tasks. Head gesture recognition is supported by motion sensing on a smart earpiece to capture head moving signals and machine learning algorithms  (K-Nearest-Neighbor (KNN) with a Dynamic Time Warping (DTW) distance measurement).
A 10-participant user study proved that HeadText could recognize 7 head gestures at an accuracy of 94.29\%. After that, the second user study presented that HeadText could achieve a maximum accuracy of 10.65 WPM and an average accuracy of 9.84 WPM for text entry. Finally, we demonstrate potential applications of HeadText in hands-free scenarios for (a). text entry of people with motor impairments, (b). private text entry, and (c). socially acceptable text entry.

\end{abstract}



\begin{teaserfigure}
  \includegraphics[width=\textwidth]{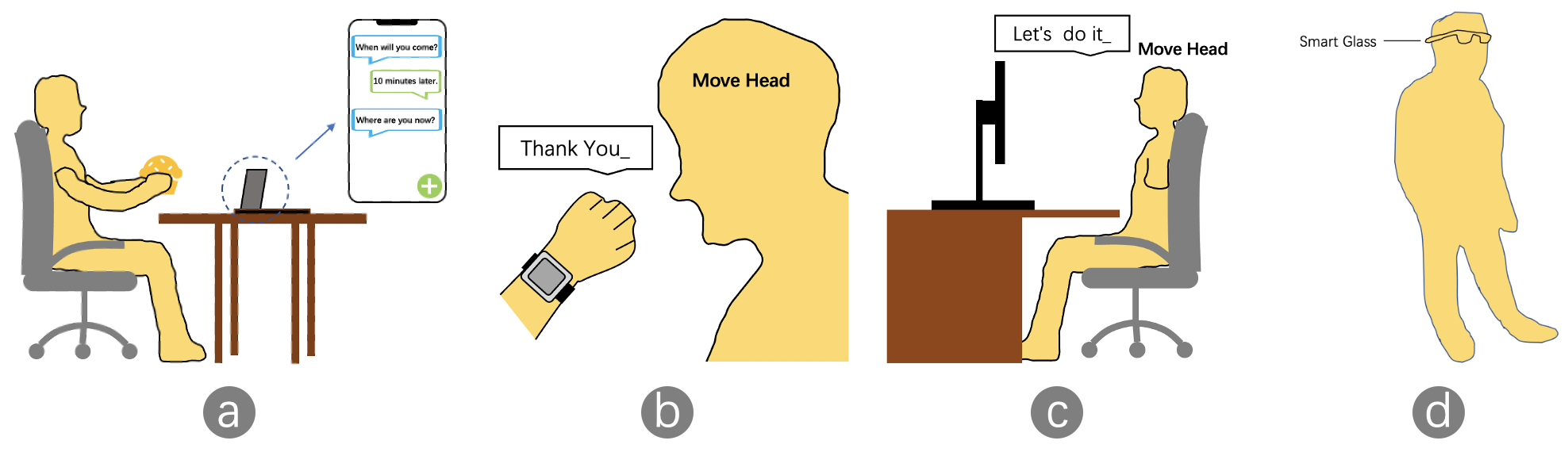}
  \caption{Application scenarios of HeadText. Fig.a shows that users could perform head gestures to respond to messages 
  when they are eating and their hands are holding the cake. Fig. b presents that a user could use HeadText to input text
  on a smart watch, whose text entry is originally limited due to the small screen size. Fig.c shows that HeadText 
  could be utilized to help users with motor impairments for text entry. Fig. d presents the application scenario where
  users could perform head gestures for text entry on the smart glass which is private and socially acceptable in public.}
  \Description{.}
  \label{f1_application_scenario}
\end{teaserfigure}

\maketitle

\section{Introduction}
Text entry is an important and common task for computing devices like computers, mobile phones and smart wearables 
in the daily life. Hands-free techniques provide a new interaction experience for text entry, which is especially useful for 
people whose hands are occupied with other tasks or people with motor impairments. Therefore, hands-free text entry has 
become an important problem to assist users for more flexible text input. 

One apparent solution to realize hands-free text entry is speech input. However, this may be inappropiate in some social 
scenarios and may also leak the privacy of users. In addition, speech recognition may not work well in the noisy environment.
To address this problem, researchers have explored various kinds of techniques for hands-free text entry.
Non-vocalized speech input such as TongueBoard\cite{10.1145/3311823.3311831} and CHANTI\cite{sporka2011chanti} is a new way 
which could support hands-free text entry while also keeping users' privacy
in social situations. What's more, facial movements could also be used for hands-free text entry. FaceInput\cite{guan2019faceinput} 
presented a hands-free and secure text entry system by facial vibration. Additionally, to achieve hands-free text entry,
Eye movements\cite{abdrabou2019calibration} are investigated a lot using gaze paths(EyeSwipe\cite{10.1145/2858036.2858335}), 
blink gesture(BlinkWrite\cite{mackenzie2011blinkwrite}) or even eyebrow movements\cite{felzer2014text}.

In this paper, we report a new technique, HeadText, for hands-free text entry as well as solving private and 
socially acceptable issues. Instead of using head pointing techniques to select all keys on a QWERTY 
keyboard\cite{yu2017tap}, HeadText allows users to perform four head gestures to trigger each key on 
a 2$\times$2 keyboard in an alphabetical order to reduce head pointing errors and also improve user 
experience so that users do not have to point their heads towards all kinds of directions in the 3D 
space. Besides, our techniques are not only suitable for Head-Mounted Devices, but also for computers, 
mobile devices and smart wearables since our method does not rely on HMDs and our head gesture recognition 
system could also work well in noisy scenarios during walking, eating or speaking. 

There are three main challenges that may prevent users from using head gestures for text entry. First, 
head movements may distract users from doing their own jobs. And users may not be able to focus 
on the keyboard if they rotate their heads with too much angle. To find a balance between recognition and 
distraction, we have conducted 
a user study to understand users' head movements and explore how slight is enough for users to perform 
head gestures so that they could focus on the keyboard while performing head gestures. Second, text entry 
information may leak in public. HeadText solved this problem by allowing subtle head gestures and unknown 
gesture-text decoding mechanism. Third, general text entry behavior may be not socially acceptable in 
public, especially for users who are talking with others face to face. HeadText overcomes this challenge 
and our study shows that HeadText is more socially acceptable for spectators compared with smart phones while users are 
talking with others.

The contributions of this paper are as follows.
\begin{itemize}
  \item A motion sensing pipeline on smart wearables to detect and recognize head gestures robustly and accurately, even in noisy scenarios including walking, eating, and speaking.
  \item A new hands-free text entry system using slight head movements on an earpiece through motion sensing.
  \item Three user studies to understand users' head movements, evaluate text entry performance, and investigate socially acceptable text entry respectively.
  \item Hands-free application scenarios of HeadText in (a). text entry of people with motor impairments, (b). private text entry, and (c). socially acceptable text entry.
\end{itemize}

\section{Related Work}
HeadText is a head gesture based text entry technique to support hands-free input. Therefore, we first introduce hands-free gesture input techniques. After that, we review existing work about gesture-based and hands-free text entry. 
Finally, we present previous research that are most related with our work and state our specific 
contributions compared with existing work.

\subsection{Hands-free Gesture Input}

Eye gesture input is the most frequently used method for hands-free input such as eye gesture recognition\cite{10.1145/2578153.2583039,10.1145/2493988.2494329,10.1007/s00779-014-0818-8} and gaze interaction (Orbits\cite{esteves2015orbits},  GazeTap\cite{Hatscher2017GazeTapTH}). And mouth-related interface (Whoosh\cite{10.1145/2971763.2971765}, TieLent\cite{10.1145/3399715.3399852}) such as tongue interface\cite{10.1145/3311823.3311831,10.1145/2702123.2702591} and teeth interface (TeethTap\cite{10.1145/3397481.3450645},Bitey\cite{10.1145/2935334.2935389}, EarSense\cite{10.1145/3372224.3419197}) also offers users a novel way for hands-free input. 
What's more, ear-based interaction (EarRumble\cite{10.1145/3411764.3445205}), waist gestures (HulaMove\cite{10.1145/3411764.3445182}) and foot gestures (FootUI\cite{10.1145/3411763.3451782}, FEETICHE\cite{10.1145/3359997.3365704}) are also explored by researchers. Compared with existing work, head gestures present another way for hands-free input like navigation in 3D space\cite{10.1145/2503713.2503754}. To recognize 
head gestures or orientations, researchers have explored various sensing techniques such as motion sensing\cite{10.1145/3197768.3201574}, 
acoustic sensing (Soundr\cite{10.1145/3313831.3376427}), capacitive sensing\cite{10.1145/2634317.2634328}, 
vision-based sensing\cite{10.1145/1111449.1111464,10.1145/1378063.1378157}, and so on. Radi-Eye\cite{10.1145/3411764.3445697} is a hands-free radial interfaces for interaction 
in 3D space using both gaze and head crossing gestures. HeadCross\cite{10.1145/3380983} and HeadGesture\cite{10.1145/3287076} presented a hands-free interface using head movements for HMDs. In addition, 
Headbang\cite{10.1145/3379503.3403538} and HeadReach\cite{10.1145/3313831.3376868} used head gestures for discrete action triggering and reachability increasement on mobile devices. 
What's more, Andrea Ferlini et al.\cite{10.1145/3345615.3361131} used in-ear wearables to track head motions, which is similar with our technique using an 
earpiece to recognize head gestures.

\subsection{Hands-Free and Gesture-Based Text Entry}

Eye movements\cite{abdrabou2019calibration} are investigated a lot for hands-free text entry, whether using gaze interaction(EyeSwipe\cite{10.1145/2858036.2858335}), blink gestures (BlinkWrite\cite{mackenzie2011blinkwrite}, BlinkWrite2\cite{ashtiani2010blinkwrite2}) or eyebrow movements\cite{felzer2014text}. Lu et al.\cite{lu2020exploration} also examined the difference between DwellType, BlinkType and NeckType in virtual reality. Additionally, non-vocalized speech input is another useful method such as TongueBoard\cite{10.1145/3311823.3311831} and CHANTI\cite{sporka2011chanti}. Naoki Kimura et al.\cite{10.1145/3411763.3451552} used a SilentSpeller to realize mobile and hands-free silent speech texting by sensing tongue gestures. Facial movements could also be used for hands-free text entry by facial vibration detection (FaceInput\cite{guan2019faceinput}) or vision (Face Typing\cite{gizatdinova2012face}). Yulia Gizatdinova et al.\cite{gizatdinova2012comparison} examined the difference between text entry using eye tracking, head tracking, and facial gestures including mouth open and brows up gestures in a video-based interface and found that mouth interaction caused significantly fewer errors than brow interaction.

Additionally, head gestures could also be utilized for hands-free text entry. Usually, head gestures are used for 
pointing and assisting key selection in AR/VR Head-Mounted Displays\cite{xu2019pointing,speicher2018selection}.
DepthText\cite{lu2019depthtext} leveraged head movements towards the depth dimension for hands-free text entry in 
mobile virtual reality systems. RingText\cite{xu2019ringtext} presented a dwell-free and hands-free text entry technique 
for mobile head-mounted displays using head motions. Chun Yu et al. \cite{yu2017tap} explored the difference of 
head-based text entry techniques using tap, dwell or gesture input method, i.e. TapType, DwellType and GestureType.
However, these head-based text entry techiques(RingText\cite{xu2019ringtext}, DepthText\cite{lu2019depthtext}, 
TapType, DwellType and GestureType\cite{yu2017tap}) all focused on Head-Mounted Displays, which may not be suitable for general 
computing devices such as computers, mobile phones, and smart wearables. What's more, these techniques used built-in IMU 
sensor in HMDs for head motion tracking, but they did not explore and evaluate the effect of users' movements such as 
walking, eating or speaking that may have an influence on their head-point techniques. Additionally, pointing each key 
of the keyboard using head motion tracking may lead to apparent errors in some specific situations such as noise scenarios.

Our work, HeadText, is designed to provide hands-free text entry using head gestures which is suitable for general computing 
devices including not only HMDs and computers, but also mobile devices and smart wearables. We employ a 2$\times$2 keyboard 
layout which is controlled by only four head gestures to reduce the errors that may be caused by head pointing and also 
provide better user experience so that users do not have to point their heads towards too many directions for key selection.
In addition, we conduct a user study to understand users' head movements and also explore the private and socially acceptable text entry challenges. Our technique is also low-cost, as it only relies 
on an IMU sensor on an earpiece, which could work with other computing devices such as smart watches, smart phones and 
computers together for text entry.

\section{Design Considerations}
\label{section 3}

\begin{figure}
  \includegraphics[width=\linewidth]{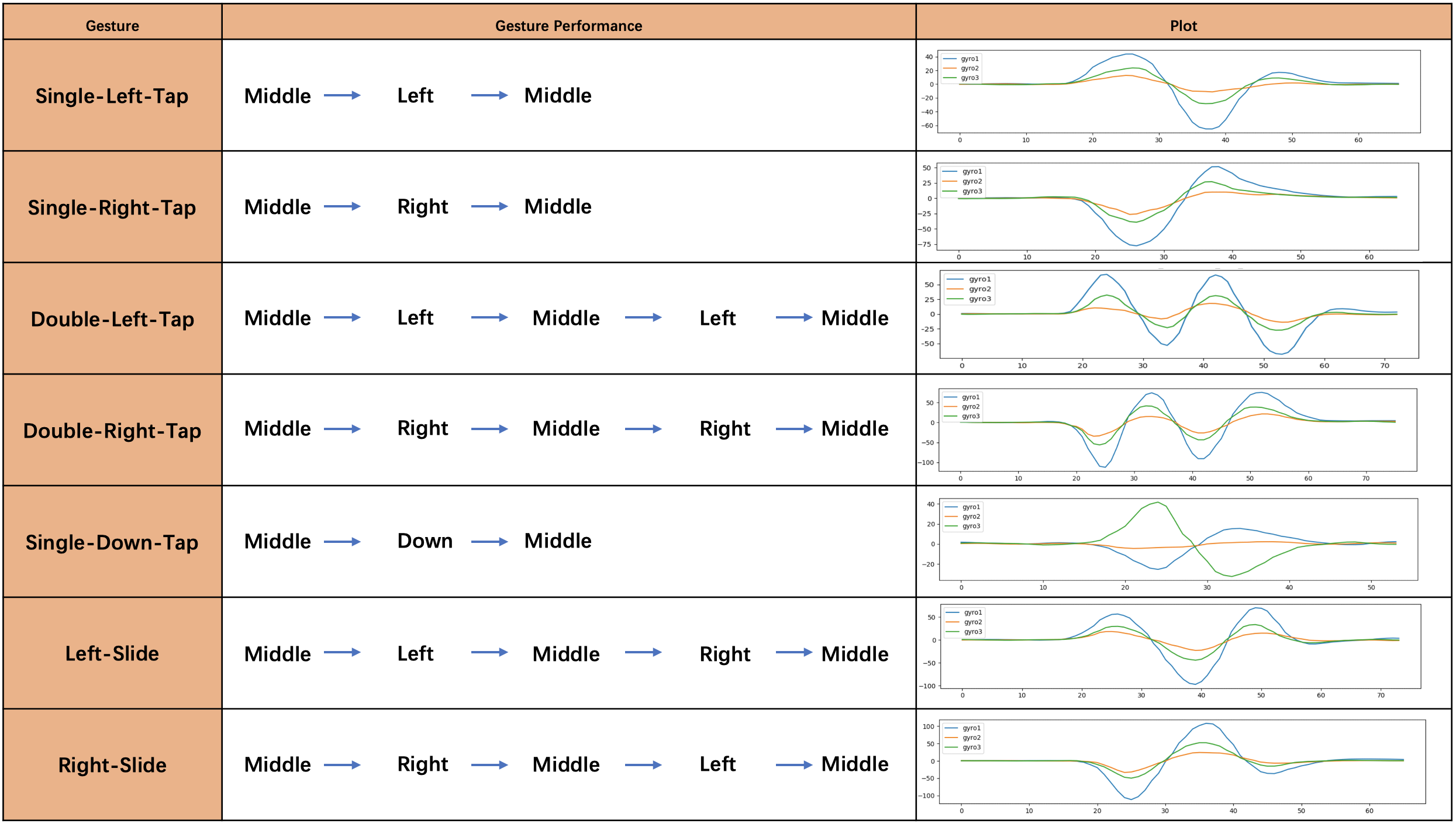}
  \caption{Head gestures sets and gyroscope signals.}
  \Description{.}
  \label{f2_head_gesture_set}
\end{figure}

We considered the following factors in the HeadText design.

\subsection{Learnability}
Learnability is important during our design process. We need to make sure that users could learn our technique
quickly and easily. There are several factors that are related to users' learnability. First, the head gesture 
set should be designed carefully. Too many or complex head gestures may be hard for users to remember and
perform. In addition, we also need to consider the design of keyboard layout, 
including the one following an alphabetical order or a QWERTY layout. Considering that some users may not 
be a master of QWERTY keyboard, we decide to adopt the alphabetical order so that most users could get used 
to our system as quickly as possible.

\subsection{User Experience}
The user experience here mainly refers to users' feelings when they are using head gestures for text entry.
Specifically, in the text entry process, some head gestures will be performed for many times. If the
gestures are too complicated, users will feel pretty tired. Besides, the design of text entry system, such as 
mapping head gestures to the commands of text entry system, should be designed carefully. For instance, the direction of head gestures could be related to the position
of keys on the keyboard. In this case, users will feel easy and intuitive to perform head gestures for text entry.

\subsection{Accuracy and Efficiency}
There are two types of accuracy: head gesture recognition accuracy ({\it $E_1$}) and text entry accuracy({\it $E_2$}). Higher {\it $E_1$} could speed up the
text entry process and result in higher {\it $E_2$}. However, the text entry system could also tolerate 
certain level of head gesture recognition errors thanks to the auto-correct model in our text entry system. 
The lasting time of head gestures will affect text entry speed, and therefore has
an impact on the efficiency of text entry process. Besides, {\it $E_1$} will
affect users' text entry performance. If a head gesture is not recognized correctly, the user has to perform
the gesture again, decreasing {\it $E_2$}. Word disambiguation appears if more than one letters 
are associated with an enlarged key (e.g. T9) which could be hard to tell which letter the user wants to enter
\cite{10.1145/3332165.3347865}. Thus, a balance is required between word disambiguation and key size.

\subsection{Privacy and Social Acceptability}
To protect users' private text entry information, head gestures should not be easily noticed. Even if head gestures are noticed, others should not be able to decode text entry information from head gestures. Socially acceptable text entry issue appears in public, especially when users are talking with others face to face. For instance, if users are talking with others while using phones for text entry at the same time, others may feel uncomfortable who may think phone users do not respect them enough. Similarly, head gestures should not be too obvious as well.

\section{HeadText: System Implementation}
\label{section 4}

\begin{figure}
  \includegraphics[width=\linewidth]{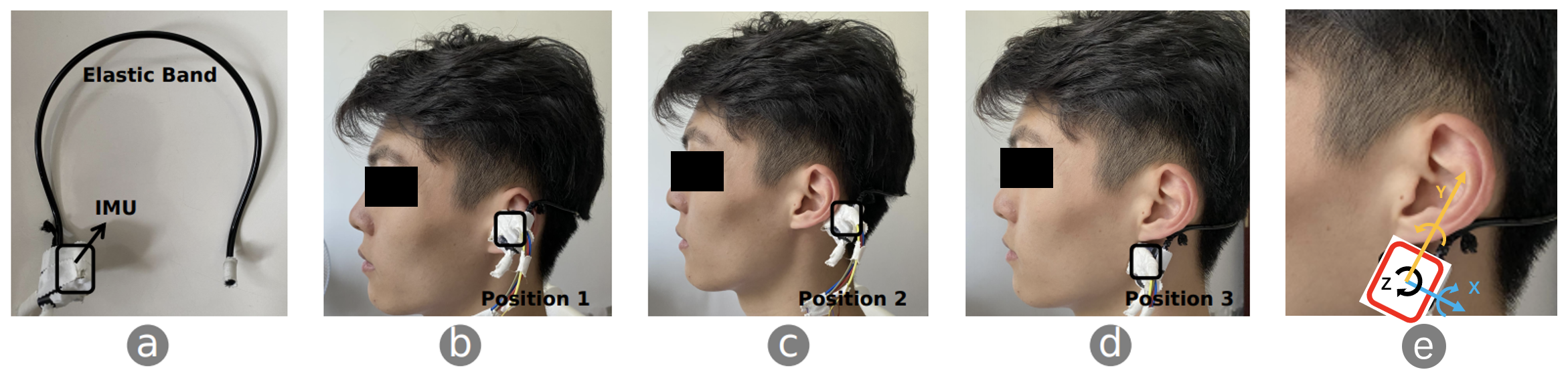}
  \caption{(a). HeadText prototype. (b)(c)(d): three sensor 
  locations. (e): 3-axis (X, Y, and Z) direction of gyroscope.}
  \Description{.}
  \label{f3_sensor_location}
\end{figure}

\subsection{Head Gesture Design}
The design 
of head gesture set is inspired by HeadGesture\cite{10.1145/3287076}. In order to make sure our gesture set is easy to be memorized
and performed by users, we choose four directions where users' heads will perform: Left, Right, Up and Down. 
To enlarge the head gesture set, we introduce different combination of head gestures towards these four directions while also considering the design considerations in Section \ref{section 3}. Finally we have designed seven head gestures which are shown in Fig. \ref{f2_head_gesture_set}.

\begin{itemize}
  \item Single-Left-Tap. Turn left from the middle position, and turn right to return to the middle position.
  \item Single-Right-Tap. Turn right from the middle position, and turn left to return to the middle position.
  \item Double-Left-Tap. Perform Single-Left-Tap gesture twice.
  \item Double-Right-Tap. Perform Single-Right-Tap gesture twice.
  \item Single-Down-Tap. Turn down from the middle position, and turn up to return to the middle position. 
  \item Left-Slide. Turn left from the middle position, and then turn right to the right side. Finally, turn 
  left to return to the middle position.
  \item Right-Slide. Turn right from the middle position, and then turn left to the left side. Finally, turn 
  right to return to the middle position.
\end{itemize}

\subsection{Hardware Prototype}

\begin{figure}
  \includegraphics[width=\linewidth]{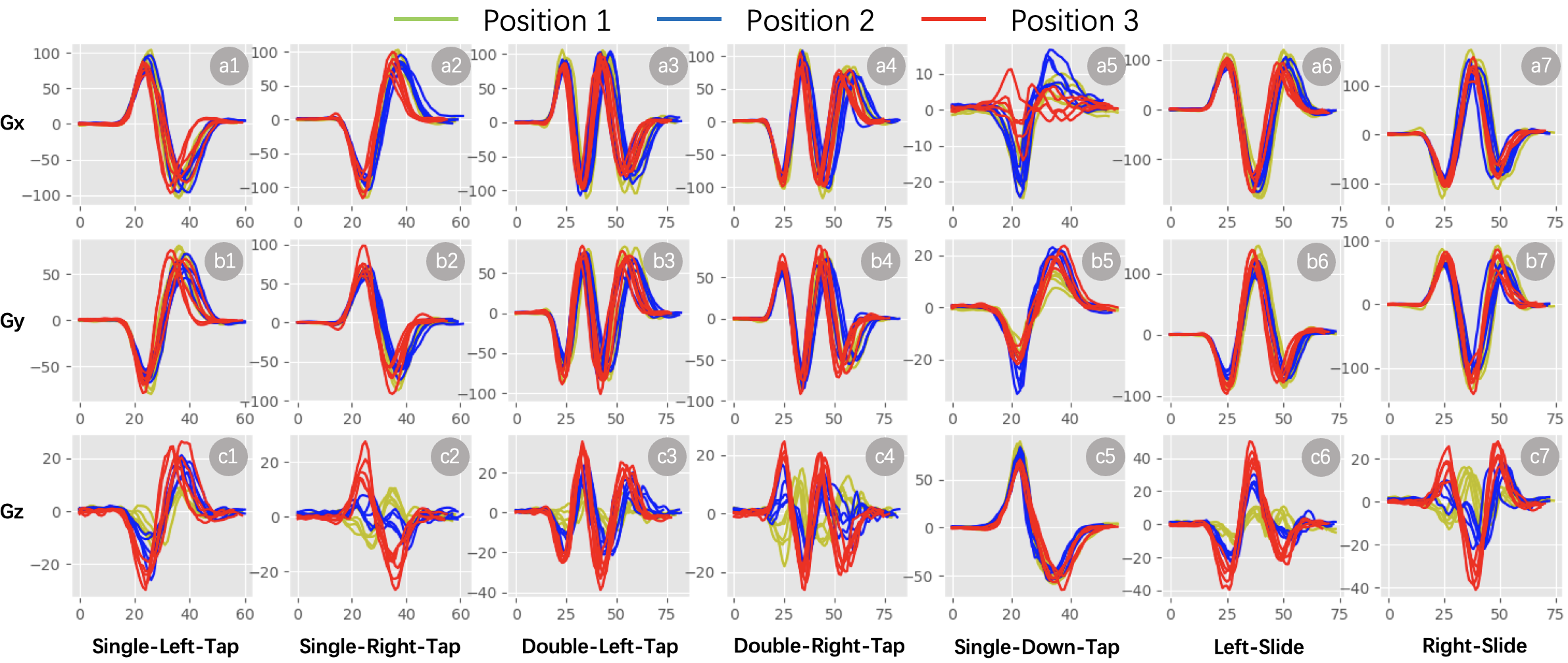}
  \caption{Gyroscope signals in three locations. Three rows stand for 3-axis of gyroscope. 
  Seven columns present signal curves of seven head gestures repeated five times randomly.}
  \Description{}
  \label{f4_sensor_loc_signal_curve}
\end{figure}

We utilize an inertial
measurement unit(IMU) to capture motion signals while users are performing head gestures. IMU is a low-cost (less than 10 dollars) and highly sensitive motion sensor. The overall
hardware system is shown in Fig. \ref{f3_sensor_location}(a). The elastic band is used to fix IMU on 
the user's head under the ear tightly enough.
While users are peforming head gestures, IMU will capture motion signals, which will be sent to the
computer to process the data through Arduino Mega. 
The IMU sensor we used contains three circuits: ITG3200, ADXL345 and HMC5883L, corresponding to
gyroscope, accelerometer, and magnetometer. After a pilot study, we finally decided to utilize gyroscope 
signals for head gesture recognition, which could capture motion signals of head gestures robustly. The direction of each axis in gyroscope is shown in Fig. \ref{f3_sensor_location}(e).
The gyroscope signal curves of seven head gestures are shown in Fig. \ref{f2_head_gesture_set}. 
In the signal curves of seven head gestures, $gyro1$, $gyro2$ and $gyro3$ stand for 3-axis of gyroscope 
data and the unit is degrees of angle per second ($degrees/s$) in the curves.

\subsection{Sensor Location Exploration}
Another problem is to decide the position where IMU is located. Theoretically speaking, IMU signal change should be similar on different positions around ear as long as IMU is fixed tightly enough, because IMU and the head constitute a rigid body and do not move relative to each other. We conducted a preliminary study to verify this assumption. Three locations are selected: on the ear canal
({\it Position 1}: Fig. \ref{f3_sensor_location}(b)), behind the ear({\it Position 2}: 
Fig. \ref{f3_sensor_location}(c)), 
and under the ear({\it Position 3}: Fig. \ref{f3_sensor_location}(d)). We asked a participant to perform these 
seven head gestures mentioned above(Fig. \ref{f2_head_gesture_set}) five times per gesture randomly in each sensor position 
scenario. Then we compared gyroscope signals which are shown in Fig. \ref{f4_sensor_loc_signal_curve}. 
We could find that all three locations have shown good
repeatability across seven kinds of head gestures, which verify our assumption.
We finally decided to put the IMU under the ear(i.e. {\it Position 3}: Fig. \ref{f3_sensor_location}(d)) 
which is easier to fix IMU and also more comfortable for users.

\subsection{Real Time Head Gesture Segmentation}

Before gesture recognition, a head gesture is first segmented from real-time series data(Fig. \ref{f5_gesture_segmentation_recognition_system}), 
which should meet requirements below:

\begin{figure}
  \includegraphics[width=\linewidth]{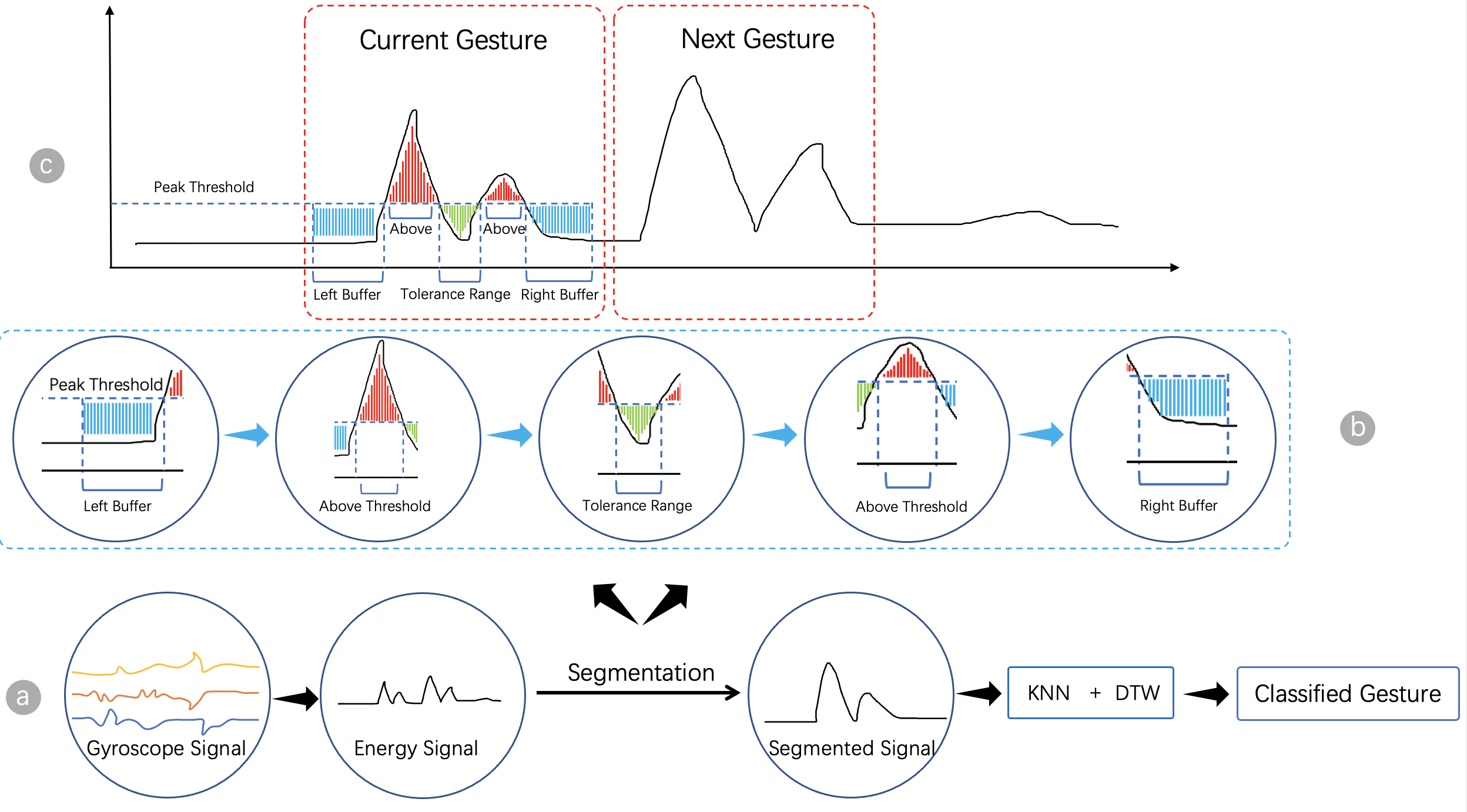}
  \caption{HeadText data processing pipeline.}
  \Description{.}
  \label{f5_gesture_segmentation_recognition_system}
\end{figure}

\begin{itemize}
  \item All head gestures 
  should be detected. Noise gestures could be 
  filtered by a noise gesture classifier later. 
  \item Head gestures should be segmented completely, instead of leaving broken gesture pieces. 
  \item Current gesture should be segmented accurately, without pieces of the next gesture or 
  the previous gesture.
\end{itemize}

To achieve this, we first calculated energy signal of the 3-axis gyroscope signals using the formula below:

\begin{equation}
  Gyroscope_{Energy}=\sqrt{Gyroscope_{x}^{2} + Gyroscope_{y}^{2} + Gyroscope_{z}^{2}} 
  \label{eq1:signal_enegry}
\end{equation}

Then we use the energy signal for gesture segmentation. As is shown in 
Fig. \ref{f5_gesture_segmentation_recognition_system}(c), there are four parameters responsible for 
gesture segmentation: {\it Peak Threshold}, {\it Left Buffer}, {\it Tolerance Range}, and {\it Right Buffer}. 
One segmented gesture is denoted as {\it Gesture Unit}. 


Fig. \ref{f5_gesture_segmentation_recognition_system}(b) shows each step in gesture segmentation 
process. 
At first, if users are keeping still, the energy will not exceed {\it Peak Threshold}, which 
will not be recorded ({\it Gesture Unit = Blank}).
After that, if the energy exceeds {\it Peak Threshold},
the system will recognize it as the start of one gesture and record the most recent 
gesture data (i.e. {\it Left Buffer}). 
At this time, {\it Gesture Unit = Left Buffer}. 
Then, if the energy is always larger than 
{\it Peak Threshold}, it will be added into {\it Gesture Unit} continuously.
After that, if the energy is lower than {\it Peak Threshold}, the system will not stop collecting data 
immediately. Instead, there is a {\it Tolerance Range}, which means that the system 
will continue saving new data to {\it Gesture Unit}, unless the lower value data length is longer than 
{\it Tolerance Range}. 
This step aims to avoid segmenting gesture data 
into broken pieces and also distinguish if a gesture has ended or not.
Finally, if energy value is lower than {\it Peak Threshold} for a longer data length than {\it Tolerance Range},
the gesture is recognized as being finished. And the {\it Right Buffer} (most recent data of 
gesture ending point) is added to {\it Gesture Unit} as the end of segmented head gesture. 

Note that although we use energy signal to segment gestures, we still use 3-axis gyroscope data for head 
gesture recognition.

\subsection{Head Gesture Recognition}
As is shown in Fig. \ref{f5_gesture_segmentation_recognition_system}(a), the segmented gesture is then 
recognized as specific head gestures using KNN (K-Nearest Neighbor) with a metric of DTW (Dynamic Time Warping).
To demonstrate the algorithm, we denote two time series head gesture data as $M_{1}^{L_{M}}$ and 
$N_{1}^{L_{N}}$. $L_{M}$ and $L_{N}$ stands for their data length respectively. $M_{i}$ and $N_{j}$ is 
denoted as the i-th point and j-th point in $M_{1}^{L_{M}}$ and $N_{1}^{L_{N}}$ 
respectively. $\delta_{i, j}$ is denoted as the distance between $M_{i}$ and $N_{j}$.
The time series data similarity measurement problem is then converted to find an optimal 
way from $\delta_{1, 1}$ to $\delta_{L_{M}, L_{N}}$. We denote $DTW_{i, j}$ as the distance between 
$M_{1}^{i}$ and $N_{1}^{j}$ with the best alignment. 
$DTW$ is first initialized as a matrix full of infinity number with $L_{M} + 1$ columns and 
$L_{N} + 1$ rows. So we have:

\begin{equation}
  DTW_{i, j} = \infty (0 < i < L_M + 1, 0 < j < L_N + 1) 
  \label{eq2}
\end{equation}

\begin{equation}
  DTW_{0, 0} = 0
  \label{eq3}
\end{equation}

For i $\in$ [$1, L_M + 1$) , j $\in$ [$1, L_N + 1$), based on dynamic planning:

\begin{equation}
  DTW_{i, j} = \delta_{i, j} + \min(DTW_{i-1, j} + DTW_{i, j-1} + DTW_{i-1, j-1}) 
  \label{eq3}
\end{equation}

Finally, the output, i.e. $DTW_{L_{M}, L_{N}}$, is the similarity measurement of 
the two time series head gesture data $M_{1}^{L_{M}}$ and $N_{1}^{L_{N}}$.

K-Nearest Neighbor(KNN) is a machine learning algorithm (implemented by scikit-learn\cite{scikit-learn}) for gesture recognition based on distance
measurement and voting. The basic idea of KNN is to find K most adjacent samples of a new sample. If 
most of the K samples belong to a certain gesture class, then the new sample also belongs to this gesture 
class. Therefore, there are four steps in our head gesture recognition algorithm:

\begin{itemize}
  \item Step 1: Data preprocessing to make sure the new gesture data could be fed into our algorithm.
  \item Step 2: Calculating the distance between the new gesture data and existing training samples 
  using DTW Measurement.
  \item Step 3: Ranking all distances from Step 2 and pick K samples with the smallest distance. 
  Here we set K = 1.
  \item Step 4: The gesture class that the new sample belongs to is that most of the K samples belong to. 
\end{itemize}

\subsection{Noise Gesture Cancellation}

\begin{figure}
  \includegraphics[width=\linewidth]{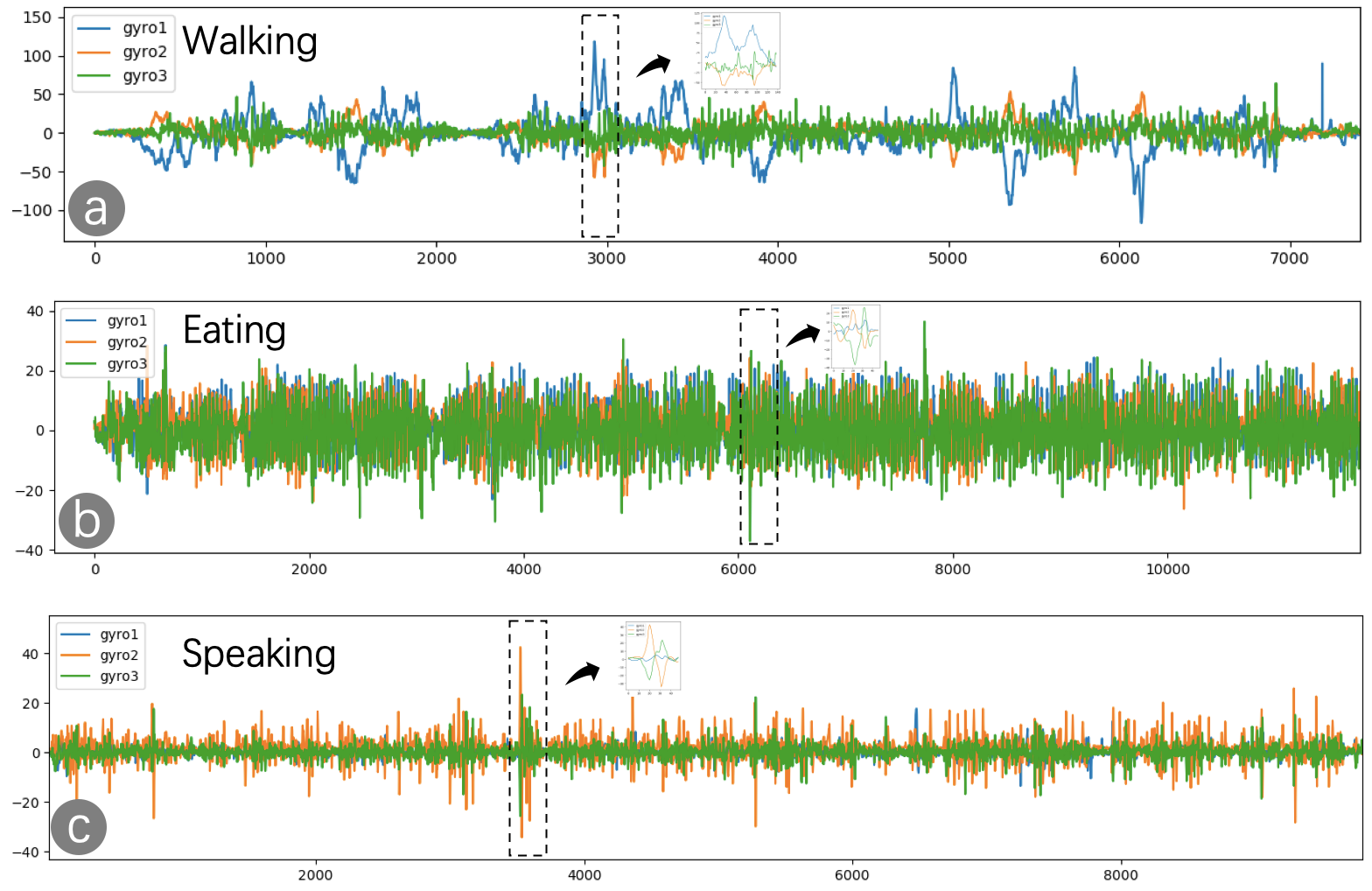}
  \caption{Gyroscope signals of walking, eating and speaking.}
  \Description{.}
  \label{f11_noise_gesture_signal_curve}
\end{figure}

We select three scenarios which are highly 
related to noise head movements and also may occur in text entry tasks: Walking, Eating and Speaking. We recruited one participant randomly and let him perform all kinds of noise gestures during these three scenarios. Each scenario lasted 2 to 3 minutes. Totally, 81 noise gesture samples were segmented including 72 walking samples, 8 eating samples and 1 speaking sample. The result also showed that compared with eating and speaking, walking will have a larger influence on head gesture recognition. This is reasonable because walking will cause larger head movements while eating and speaking will not have such affect apparently. The noise data from three scenarios are shown in Fig. \ref{f11_noise_gesture_signal_curve} and we could find that there are indeed a list of gesture samples that may be confused with head gestures. 
However, we find that time domain features of energy curves such as peak number is clearly different between noise gestures and head gestures. 
This provided us theoritical proof to train a binary classifier to distinguish them using 
energy signals. We also asked the participant to perform each head gesture for three times as the standard head gesture data set. Finally, we used energy signals of 81 noise samples and 
21 head gesture samples for training (80\% data) and testing (20\% data) a binary SVM classifier. 
The testing result indicated that all gestures are classified correctly. This classifier also 
worked well in our following user study for head gesture recognition 
and text entry tasks.

\section{Understanding Head Movements}
In the text entry system, speed of head movements is a trade-off problem.
On one hand, high speed head motions could result in high text entry speed. However, on the other hand, 
high speed head motions may generate noisy head gestures, result in decreasement of head gesture recognition accuracy. 
Additionally, high speed head motions may lead to uncomfortable user experience, distracting users from focusing on
the keyboard layout and even cause private and socially acceptable issues. Therefore, we conduct the first 
user study to understand users' head movements and find the optimal {\it Peak Threshold} value, which is directly related 
to head motion speed (described in Section \ref{section 4}.4).
The initial range of {\it Peak Threshold} is from 10 to 50 $degrees/s$, which is obtained 
from a pilot study. Then we select five value: 10, 20, 30, 40, and 50 for exploration. 
To find the best suitable value, we implement each one in each session and analyze head gesture recognition 
result in the study. 

\subsection{Participants}
We recruited 10 right-handed participants (4 female) aged between 20 and 26 from the author's institution. 
Each participant was asked to fill in a questionnaire and was paid with 10 USD for their time.

\begin{figure}
  \includegraphics[width=\linewidth]{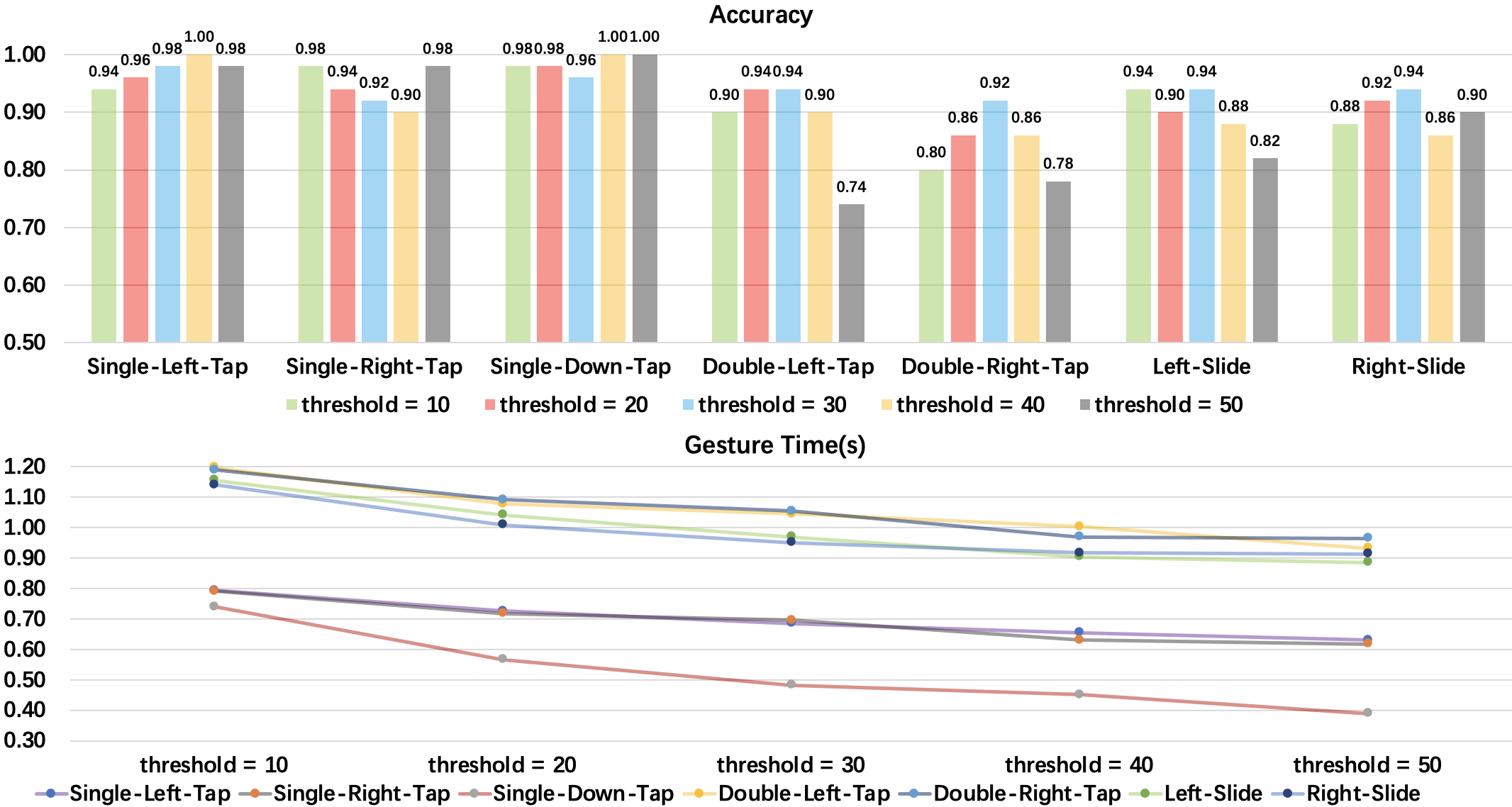}
  \caption{Head gesture recognition accuracy 
  and average lasting time of each gesture under different {\it Peak Threshold}.}
  \Description{.}
  \label{f8_u1_result_accuracy_gesture_time}
\end{figure}

\subsection{Apparatus}
We utilized the prototype of HeadText which was described before in Fig. \ref{f3_sensor_location}(a) 
to conduct the user study. A computer screen was in front of each participant to provide visual instructions 
such as which kind of head gesture that the participant will do to the participant. 

\subsection{Study Design}

There are five sessions for each participant totally, corresponding to five value settings of 
{\it Peak Threshold}. 
Participants do not know the different {\it Peak Threshold} value set. Also, the order of different 
sessions is shuffled randomly.
In each session, there are two parts: training part and testing part. 
During each training part, participants need to perform 3 samples per head gesture randomly 
to train a head gesture recognition model, which will be used to recognize head gestures in 
the testing part, where participants are asked to perform 5 samples per head gesture randomly.
Totally, for the ten participants, we collect 10$\times$5$\times$(3+5)$\times$7=2800 gesture samples.

\subsection{Procedure}
The researchers first introduced the study to participants and then instructed them to wear the prototype 
and explained visual instructions 
on the computer. Each participant was allowed to practice head gestures before they 
conducted actual tests until they felt confident to perform head gestures correctly. 
For each session, participants followed instructions on the screen to 
perform head gestures. If they happened to make mistakes such as wrong gestures, they would 
tell experimenters to remove wrong samples. The groundtruth was recorded by the computer 
automatically. Participants did not know whether performed gestures were recognized correctly. 
The whole user study took about 1 hour for each participant, who would fill in a NASA TLX 
form after each session.

\subsection{Result and Analysis}

\begin{figure}
  \includegraphics[width=\linewidth]{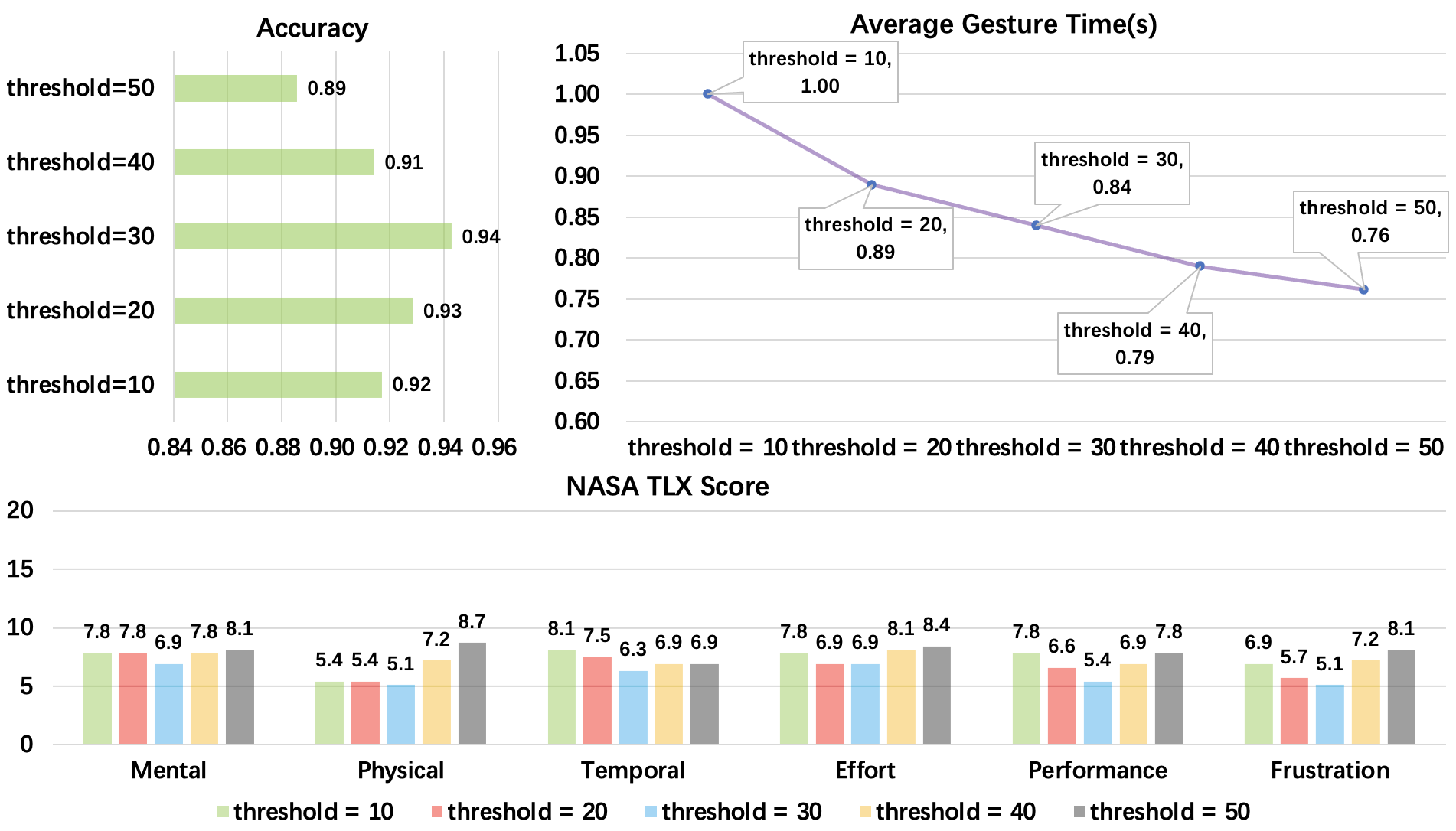}
  \caption{The first user study results.}
  \Description{.}
  \label{f9_u1_result_average_accuracy_time_NASA}
\end{figure}

\subsubsection{Head Gesture Recognition Accuracy}
Fig. \ref{f8_u1_result_accuracy_gesture_time} shows recognition accuracy 
during different sessions(i.e. different {\it Peak Threshold}). We could find that most results 
are higher than 90\%. However, compared with others, threshold = 30 performs best in 
most head gestures such as {\it Double-Left-Tap}, {\it Double-Right-Tap}, {\it Left-Slide} and 
{\it Right-Slide}. Although threshold = 30 performs worst in {\it Single-Down-Tap}, the recognition accuracy
(96\%) is still higher than 95\%. Besides, threshold = 30 is the only threshold that 
accuracy of all gestures is higher than 90\%. Additionally, as is shown in  
Fig. \ref{f9_u1_result_average_accuracy_time_NASA}, threshold = 30 holds the highest gesture
recognition average accuracy (i.e. 94\%). 

\subsubsection{Head Gesture Lasting Time}
The gesture lasting time means the time from when the gesture is detected to when the gesture is recognized to be 
finished in our system, which is shown in 
Fig. \ref{f8_u1_result_accuracy_gesture_time}. We find that gesture 
lasting time will decrease with the increasement of threshold, which indicates that threshold = 50
holds the shortest gesture lasting time. However, in Fig. \ref{f8_u1_result_accuracy_gesture_time}, 
we also find that the gap of lasting time between 
threshold = 30 and threshold = 50 is actually much lower than that between threshold = 10 and threshold = 50.
In addition, as is shown in Fig. \ref{f9_u1_result_average_accuracy_time_NASA}, the gap of average gesture 
time between threshold = 10 and threshold = 30 is 0.16s while that between threshold = 30 and threshold = 50 is only 0.08s. 
If taking both high gesture recognition accuracy and low gesture lasting time into considerations, 
threshold = 30 could be the best value.

\subsubsection{Subjective Feedback}
NASA TLX result is shown in Fig. \ref{f9_u1_result_average_accuracy_time_NASA}. Surprisingly but reasonably, we found that threshold = 30 held the 
lowest score in all classes of NASA TLX form, indicating that threshold = 30 could provide the best 
user experience among all threshold. We got reasons after interviewing some participants. 
In the threshold = 10 or threshold = 20 sessions, users felt that some noise gestures 
were easy to be detected, distracting their focus 
on performing head gestures. While for the threshold = 40 or threshold = 50 sessions, it might took 
more efforts of them to perform head gestures so that gestures could be detected by the system. 

Finally, considering the trade-off problem among gesture recognition accuracy, gesture lasting time, and 
subjective feedback from participants, which may affect both text entry speed and user experience, 
we decide to use 30 $degrees/s$ as the {\it Peak Threshold} to provide fast yet also user-friendly 
text entry system.

\section{Text Entry System}

\begin{figure}
  \includegraphics[width=\linewidth]{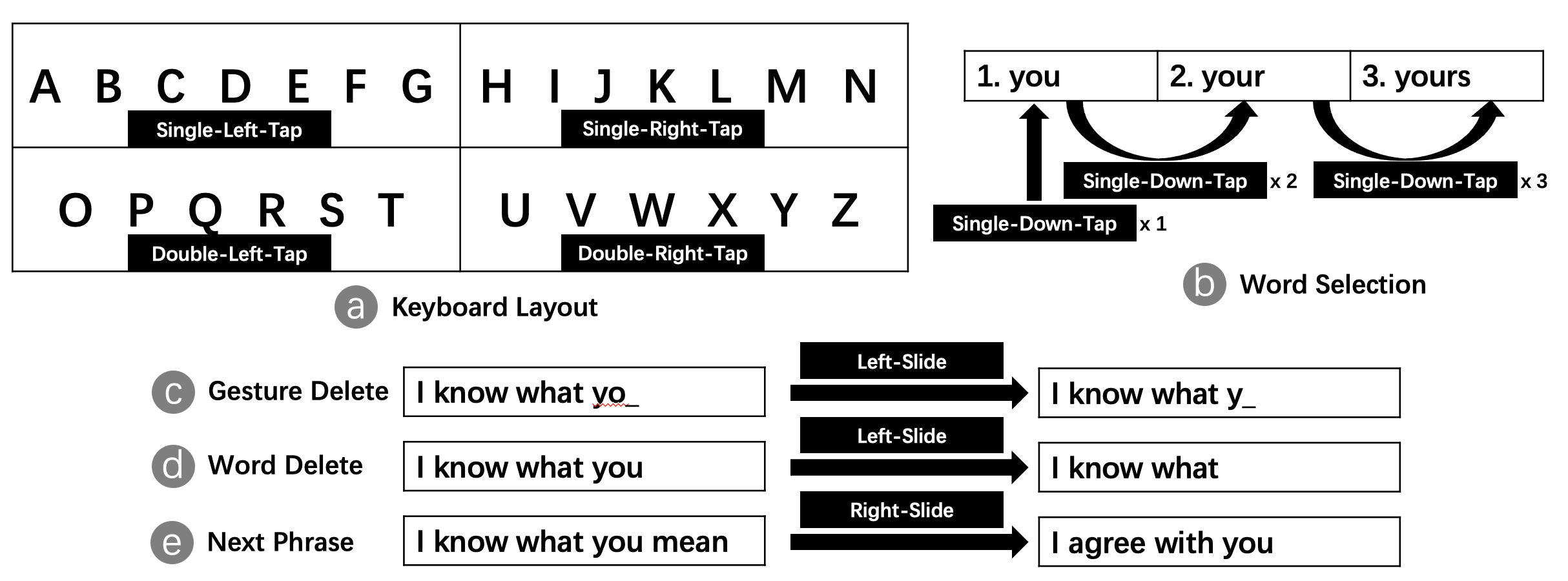}
  \caption{HeadText text entry system.}
  \Description{.}
  \label{f7_text_entry_system}
\end{figure}

\subsection{Keyboard Layout Mapping}

With design considerations in section \ref{section 3} and user study results, we show our text entry system in Fig. 
\ref{f7_text_entry_system}. We adopt the alphabetical order of the keyboard layout 
so that most users could get used to it as soon as possible. The whole keyboard is divided into four blocks, 
with each block is mapped to a head gesture. For each word to input, users could trigger specific blocks by 
performing the corresponding head gestures. Then an auto-complete algorithm will 
produce three predicted words 
according to the blocks that are triggered by users, so that users do not have to input 
all letters. Users could select one word in turn from the 
three auto-complete words by performing {\it Single-Down-Tap}. 
We choose {\it Single-Down-Tap} for word selection in a reason 
that {\it Single-Down-Tap} holds the shortest lasting time compared with other gestures (results are from 
the first user study).

The word will be committed automatically once the user starts 
typing the next word. A space will be inserted after the committed word automatically.
We also have an 
auto-correct algorithm that could predict right words even if users make a mistake about their 
head gestures. Users could also cancel the last head gesture or the last 
word they have selected by performing {\it Left-Slide}. The difference between gesture 
cancellation(Fig. \ref{f7_text_entry_system}(c)) and word cancellation
(Fig. \ref{f7_text_entry_system}(d)) is whether users have selected a word by {\it Single-Down-Tap}. 
This means if users have already commited a word by {\it Single-Down-Tap}, then when they perform 
{\it Left-Slide}, the system will delete the last word immediately. 
Otherwise, if users have not 
selected a word by {\it Single-Down-Tap}, then the system will 
delete the last gesture. After users have finished the current 
phrase, they could perform {\it Right-Slide} to start to input the next phrase(Fig. \ref{f7_text_entry_system}(e)).

\subsection{Auto-Complete and Auto-Correct}

Our auto-complete and auto-correct feature is based on WrisText\cite{10.1145/3173574.3173755}. A Bayesian model is utilized to predict a target word(denoted as $W$) based on users'
gesture input $S = [s_1, s_2, s_3, \dots, s_n]$(i.e. a series of head gestures). We denote the word dictionary we use 
as $L$. We need to find a word $W^{*}$ in $L$ that could satisfy: 

\begin{equation}
  W^{*} = \arg\max_{W \in L} P(W|S)
  \label{eq5}
\end{equation}

Additionally, according to the Bayes' rule, we have:

\begin{equation}
  P(W|S) \times P(S) = P(S|W) \times P(W)
  \label{eq6}
\end{equation}

Therefore, 

\begin{equation}
  P(W|S) = \frac{P(S|W)P(W)}{P(S)}
  \label{eq7}
\end{equation}

Because $P(S)$ is an invariant across words, we have:

\begin{equation}
  P(W|S) \propto P(S|W) \times P(W)
  \label{eq8}
\end{equation}

Therefore, Equation \ref{eq5} could be converted to:

\begin{equation}
  W^{*} = \arg\max_{W \in L} P(S|W) \times P(W) 
  \label{eq9}
\end{equation}

In Equation \ref{eq9}, $P(W)$ could be got from language model (LM) and $P(S|W)$ could be obtained from 
spatial model (SM), so we have:

\begin{equation}
  P(S|W) = \prod_{i=1}^{n} P(S_{i}|W_{i}) \times \alpha^{m-n}
  \label{eq10}
\end{equation}

We assume that users will not generate insertion or omission errors. Each key strike that is 
triggered by users' head gestures is treat independently\cite{10.1145/3173574.3173755}. In Equation \ref{eq10}, $S_{i}$ 
stands for the $i$-th letter of the word which is entered by users and $W_{i}$ refers to the $i$-th letter 
of $W$ in the dictionary $L$ whose length is between $S$ and $S+8$, where 8 is determined from both 
WrisText\cite{10.1145/3173574.3173755} and our preliminary test. To prevent long words with high frequency to be 
ranked too high in our system and may dominate short words with low frequency, we introduce the penalty 
parameter $\alpha_{m-n} (m \geq n)$, where $m$ is the length of word $W$ and $n$ is the number of head gestures 
in $S$. We finally set $\alpha = 0.65$ which performs best in our text entry system. 

Auto-correct feature is mainly used to deal with cases where users fail to select a desired key, which is 
composed of two scenarios: 1. Our head gesture recognition 
system fail to recognize users' right gestures; 2. users perform a wrong head gesture which is actually 
not the gesture they intend to perform. From the first user study we find that the 
head gesture recognition accuracy is 94\%. Therefore, we set $P(S_i|W_i) = 0.94$ if $S_i$ and $W_i$ are from 
the same key. For $S_i$ and $W_i$ belonging to keys that are adjacent or diagonal to each 
other, we choose 0.025 and 0.01 respectively.

\section{Text Entry Evaluation}

\subsection{Participants}
We recruited 5 right-handed participants (2 female) aged between 22 and 26 from the author's institution(Covid-19 has limited further recruitment). Each participant was paid with 10 USD for their time.

\subsection{Apparatus}
We utilized prototype in Fig. \ref{f3_sensor_location}(a) 
to conduct the study. A computer screen was in front of each participant to provide visual instructions such as which phrase for input to the participant. Also, top three candidates by auto-complete will be shown and participants could perform {\it Single-Down-Tap} to navigate the candidate list and commit words. Participants could also perform {\it Left-Slide} to delete the last letter or the last word. The keyboard layout was also shown on the screen to assist participants for text entry.

\subsection{Study Design}
There are six sessions in this study, each containing 8 phrases picked from the MacKenzie's phrase set\cite{10.1145/765891.765971} randomly. Each participant will input all phrases in each session and no phrase was repeated. Since different sessions contained different phrases, the order of sessions for each participant was also shuffled randomly. After entering a phrase, a participant could perform {\it Right-Slide} to proceed to the next phrase. 
Totally, we collected 5 participants $\times$ 6 sessions $\times$ 8 phrases = 240 phrases.

\subsection{Procedure}
The researchers would first introduce the study to participants and instruct them to wear our prototype and explain the visual 
instructions on the computer screen. Each participant was allowed to practice performing head 
gestures for text entry before they conducted the actual test. When participants got familiar 
with our system, they would proceed to the following six sessions. 
The groundtruth of target phrases, participants' actual input and text entry speed was recorded 
by the computer automatically. The user study took about 1 hour for each participant.

\subsection{Result and Analysis}

\begin{figure}
  \includegraphics[width=\linewidth]{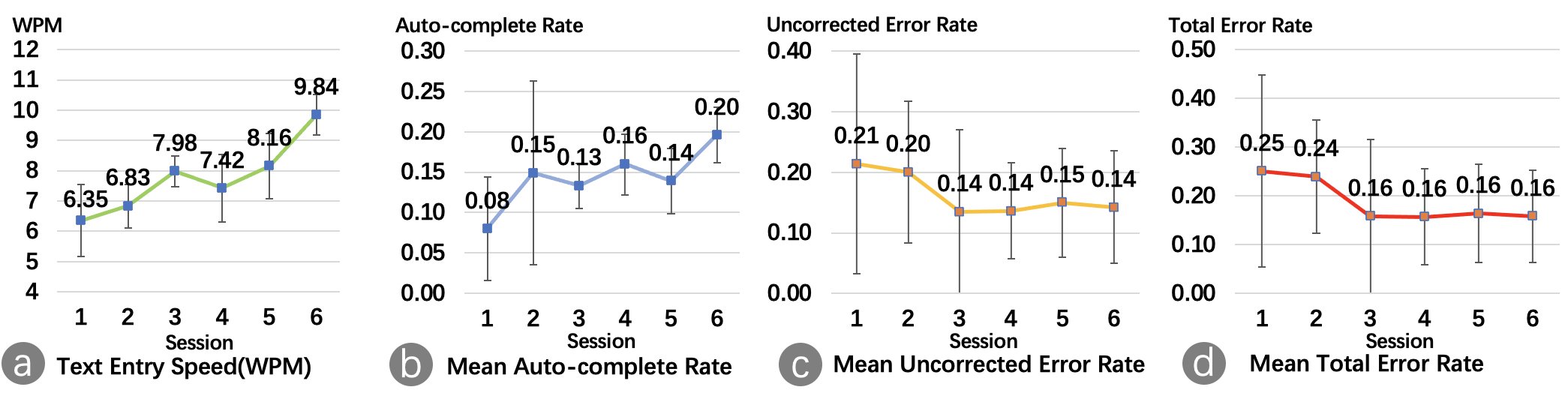}
  \caption{Text entry results.}
  \Description{.}
  \label{f10_u2_result_1}
\end{figure}

We analyzed the results through a one-way repeated measures ANOVA. Bonferroni corrections were 
used for pair-wise comparisons and a Greenhouse-Geisser adjustment was adopted for degrees of 
freedom for violations to sphericity.

\subsubsection{Text Entry Speed}
Mean WPM of each session was shown in Fig. \ref{f10_u2_result_1}(a).
ANOVA yielded a significant effect of Session (F(2.106, 8.424) = 20.456, p < 0.001). 
Fig. \ref{f10_u2_result_1}(a) also demonstrated a performance improvement through users' practice. In addition, 
Post-hoc pair-wise comparisons showed a significant difference between 
first and fifth session(p < 0.001), and also difference between first and sixth session(p < 0.05), second and sixth session(p < 0.05). 
Participants achieved 6.35 WPM (s.e. = 0.430) in the first session and the speed increased to 9.84 WPM (s.e. = 0.240) in the last session with an improvement of 55\%. 
These results suggested potentials to continue increasing text entry speed for long-term learning. 
Additionally, the maximum speed in the last session is 10.65 WPM.

\subsubsection{Error Rate}
Two kinds of error rate are reported: uncorrected error rate (UER) and total error rate (TER). Uncorrected errors refer to errors that are found in the final input words after auto-correction. Total errors include both uncorrected errors and errors that have been corrected by the system. Overall, for both UER and TER, no significant effect was found in Session
(UER: F(1.925, 7.700) = 1.864, p > 0.05; TER: F(2.040, 8.161) = 2.517, p > 0.05). Whereas UER and TER 
decreases with the increasement in text entry speed, suggesting that users increase text entry speed by decreasing mistakes. However, correcting errors was still one of the major sources preventing users from achieving faster speed. 
The average UER and TER was 16.28\% and 18.78\% respectively, indicating that our auto-correction 
feature was not fully utilized. Fig. \ref{f10_u2_result_1}(c)(d) showed results, where 
we found that average UER and TER in the first session(21.40\%, 25.10\%) improved by 34\% and 37\% in the last session(14.20\%, 15.80\%). Although error rate improvement is significant, we could find that there is still enough space to improve the error rate.


\subsubsection{Auto-Complete Rate}
The auto-complete rate of a word is calculated by dividing the number of automatically filled letters by 
the length of the word. Results were shown in Fig. \ref{f10_u2_result_1}(b).
No significant effect was found in Session (F(1.563, 6.253) = 3.075, p > 0.05). 
Participants achieved 8.00\% auto-complete rate (s.e. = 2.3\%) in the first session 
which increased to 19.60\% (s.e. = 1.2\%) in the last session with an improvement of 145\%.
Additionally, we could find that text entry speed
increases with the increasement in auto-complete rate.
These results indicated that users accelerated their text entry speed by increasing their 
auto-complete rate.

\subsubsection{Subjective Feedback}
We asked participants to 
express their opinions and suggestions. All participants agreed with our idea of using head gestures 
for text entry and some of them purposed that this technique could be used in the commercial products like smart 
earbuds. One participant said that HeadText is pretty cool and convenient for use.
Another participant hoped that we could built a more flexible band to wear more comfortably. Overall, 
participants welcomed our technique and believed that this could have many potential applications in the
future.

\section{Application Scenarios}
In addition to hands-free applications in Fig. \ref{f1_application_scenario},
we demonstrate application scenarios in three points: text entry of people with 
disabilities, socially acceptable text entry and private text input.

\subsection{Text Entry for People with Disabilities}
The most direct application of this hands-free text entry technique is to help people with motor 
impairments for text input. 
Users could perform head gestures to trigger keys on a keyboard and commit the word to the system 
instead of using hands. Compared with HeadText, GazeSpeak\cite{10.1145/3025453.3025790} is another alternative technique for 
people with Amyotrophic Laternal Sclerosis (ALS) and other motor impairments by allowing users to use gaze directions for text input 
at an average speed of 80.9 seconds per sentence (around 3.71 WPM) and at a maximum speed of 51 seconds 
per sentence (around 5.88 WPM). Whereas our evaluation shows that users mean text entry speed is 9.84 WPM and maximum 
speed is 10.65 WPM, which is much faster than GazeSpeak. Additionally, HeadText is also able to be operated 
by people with ALS, Muscular Dystrophy and other motor impairments except for users whose heads could not move.

\subsection{Socially Acceptable Text Entry}




We conducted an extra study to demonstrate social acceptability for HeadText.
Inspired by the survey of social acceptability in HCI\cite{10.1145/3313831.3376162}, we utilized two 
dimensions for evaluation: (1). the user's social acceptance and (2). the spectator's social acceptance, 
according to Montero et al.\cite{10.1145/1851600.1851647}. There are two social situations that we consider: 
face-to-face conversation (one-to-one) and public situation (one-to-more). Theoretically we needed to connect 
the prototype to a computer and ask users to 
perform gestures for real text entry while running our algorithm. However, this requires a screen 
for visual feedback of auto-complete words and word selection, which is bulky and inconvenient and may have a bias for study results.
We also anticipate that HeadText could be used with a speaker inside an earpiece or smart glasses to provide audio or visual feedback in the 
future. 
Therefore, we finally decided that HeadText did not have to be connected with a computer. Users just needed to wear 
the prototype and perform head gestures to simulate text entry.
Because our purpose is to evaluate social acceptability for wearing the prototype and performing head gestures, this design 
could also meet our research target.

\subsubsection{Face-to-face Conversation Study}

\begin{table}[]
  \begin{tabular}{|c|c|c|c|c|}
  \hline
           & \multicolumn{2}{c|}{User Acceptability} & \multicolumn{2}{c|}{Spectator Acceptability} \\ \hline
           & Wear/Hold        & Type         & Wear/Hold           & Type           \\ \hline
  HeadText & 5.67        & 5           & 5.67           & 5.67          \\ \hline
  Phone    & 7           & 5.67        & 7              & 5.33          \\ \hline
  \end{tabular}
  \caption{Social Acceptability Score in Face-to-face Conversation Scenario Wearing HeadText or Holding a Phone}
  \label{face_to_face_conversation}
\end{table}

We recruited 6 participants for the face-to-face conversation study of two sessions, who are divided into 3 pairs randomly. 
For each session, each pair was asked to have a 20-minute conversation. Participants 
could talk anything they want. The difference of two sessions is the additional task: {\it Wearing HeadText} or {\it Holding Phones} for typing, which design is 
inspired by Logas et al.\cite{10.1145/3458709.3458942}.
For each pair, we randomly selected one participant (technique user) and asked him/her to wear 
HeadText (session 1) or hold a smart phone (session 2). The user was also required to perform 20 head gestures naturally during the conversation to simulate 
inputing one phrase (around 5 words, 20 gestures) (session 1) or use fingers to input one phrase on a phone (session 2). 
Since we could not anticipate which words users want to input in the 
real life, the kind and order of head gestures or the content of words are not limited. Another participant (technique spectator) in one pair 
was not informed about tasks of the technique user. After the study, technique users were asked to score their 
acceptability to (a). wear HeadText/hold phones and (b). use head gestures/phones for typing during a conversation. Technique spectators were asked 
to score their acceptability for people who (a). wear HeadText/hold phones and (b). use head gestures/phones for typing while 
talking with them. Both technique users and spectators were encouraged to state their subjective feelings. 
The score of acceptability is presented through 1 to 7 rating. Score 1 means totally unacceptable while score 7 means 
totally acceptable. 
The result is shown in Table \ref{face_to_face_conversation}. 
We found that holding a phone owned higher acceptability score than wearing HeadText during a conversation, which was 
quite normal since users are more familiar with phones and therefore feel more natural to use a phone.
However, the spectator owns more acceptability
for people who wear HeadText {\it for typing} than those who use a phone. One participant told us that 
phone typing requires users to keep their eyes pointing at phones, instead 
of conversation partners. Therefore, spectators might not feel respected. However, using an earpiece for 
typing could make sure users' eyes focusing on spectators, even if their heads are moving at an unobservable angle.
Besides, all HeadText wearers showed their acceptability for wearing the prototype and they told us that it just 
felt like wearing a normal earphone.

\subsubsection{Public Situation Study}
The public situation (one-to-more) study is a semi-controlled study, where we designed two parts, focusing on evaluation of HeadText users and
spectators respectively. For the first part, we recruited three participants who is asked to do whatever 
they originally should do respectively for 30 minutes in three places: basketball court, library, and dining hall. For instance, a participant 
in the library may read books, talk with others and so on. In the same time, each participant needed to wear HeadText 
prototype and perform 20 head gestures. Similar with face-to-face study, users' performing gestures meet {\it three conditions}:
(a). Which head gesture the participant is performed is not limited; (b). When the participant will perform each gesture 
is not limited; (c). The spectator (others who may notice the participant in public) is not informed of the participant's 
tasks (i.e. wear and perform head gestures). 
After the study, participants were asked to state their personal feelings. The participant in the dining hall told us that he 
did not feel embarrassed because nobody noticed him except one person who swept the floor. The participant 
in the library gave us similar feedback, stating that everyone was doing their own work and therefore he did not 
have any uncomfortable feelings, just like usual work in the library. Interestingly, the participant in the 
basketball court said: " When I was performing head gestures, I felt that other observers thought that I was 
just doing some muscle and bone exercises. So I didn't feel embarrassed ". 

For the second part, we recruited one participant from the school and asked him to go to a cafeteria and do whatever he wants naturally.
Additionally, the participant needed to wear HeadText prototype and perform 20 head gestures, 
whose performing gestures meet the {\it three conditions} described before.
Overall, the participant bought a cup of coffee, played his phone and dealed with affairs work. 
There were totally 6 persons in the cafeteria except for the participant and including one salesman. We interviewed 
all the 6 spectators. Surprisingly, all spectators, including 
the salesman who sold the coffee to the participant, failed to notice that the participant was either wearing a 
prototype or performing head gestures. After an interview, the salesman told us that he was focusing on his work and therefore 
ignored the HeadText user. However, he said he might notice the user if he was not busy. When we asked his acceptability for 
the HeadText user, he said that he was tolerant of every guest's behavior. When we interviewed other spectators, they told us 
that they all did not notice the HeadText user. Also, they did not care whether users wore 
the prototype. When we interviewed the HeadText user, he told us that he felt a little embarrassed at first. But when he 
found that almost nobody noticed him and started to do his own work, he did not feel embarrassed at all.

\subsection{Private Text Entry}
Existing work (PrivateTalk\cite{10.1145/3332165.3347950}, FingerPad\cite{10.1145/2501988.2502016}, ThumbRing\cite{10.1145/2957265.2961859}) 
on private input rarely has an evaluation study for their private feature. HeadText demonstrated feasibility of 
privacy-protection through a social study in both face-to-face conversation and public situations 
(described in {\it Socially Acceptable Text Entry} part). 
The results show that most people in public (such as cafeteria, library, dining hall) will not notice the one who wear HeadText. 
The most possible situation that others may notice users' head movements is a face-to-face conversation. 
However, results also present that conversation partners rarely notice users' head motions especially if they personally do not know 
users are using head gestures for typing. What's more, even if others have noticed users' head motions, it is still hard 
to leak text entry information, because it is hard to decode text information from head gestures. Here are the reasons.

\begin{itemize}
  \item Our head gesture set is composed of four primitives: Left, Right, Up and Down. Observers may notice 
  the primitives performed by users, but they could not infer which gesture the user is performing. For instance, if 
  observers find that users are performing Left-Right-Left-Right, perhaps users are performing {\it Double-Left-Tap} or 
  two {\it Single-Left-Tap} gestures. But observers do not know which one the user is performing.
  \item It is hard for observers to remember all head gestures performed by users. If one head gesture is missed by the 
  observer, the head gesture order is interrupted and observers could not restore the text information.
  \item Observers do not know the mapping from head gestures to keyboard layout. They also do not know the word 
  prediction candidates. Therefore, even if observers know users are selecting words using head gestures, they do not 
  know which word the user has selected.
\end{itemize}

\section{Discussion and Future Work}
Although we have demonstrated the feasibility and robustness of our technique, we indeed admit that there are 
some limitations of this work and we could do more to further improve our technique.

\subsection{Text Entry Speed}
HeadText is a gesture-based system, whose efficiency highly relys 
on gesture detection and recognition time. 
Therefore, one method for increasing text entry speed is to 
speed up gesture segmentation and recognition process. One way is that the system does not have to wait until 
users have already finished a gesture. We could recognize the head gesture
using only the first half of the gesture to save time. Another way is that users do not have to turn their 
head back to the initial position after each gesture. Instead, they could directly perform the next 
head gesture. Last but not least, we could design 
faster and easier head gestures to speed up the text entry process.

\subsection{Error Rate and Auto-Complete Rate}
As is described in the result of the second user study, our auto-correct feature has not been fully 
utilized which could have better text entry performance. In our future work, we will optimize the 
parameter design in our auto-correct feature to help improve our text entry performance. We also notice 
that auto-complete rate increases with the increasement of text entry speed, indicating that users 
speed up their text entry process by utilizing the auto-complete feature. Additionally, a long-term 
user study may have higher auto-complete rate and beffer text entry performance, which could be our 
future work for exploration.

\subsection{Head Gesture Design and Mapping}
Our head gesture design is inspired by both previous work\cite{10.1145/3287076} and our preliminary study. 
However, in the studies, we find that some gestures may bring extra burden to users and also 
 may last more time which may affect text entry speed. Therefore, in our future work, 
we could further optimize our head gesture design and introduce more user-friendly and efficient head gestures 
for text entry. We will also optimize mapping specific gestures to the keys on the keyboard layout to 
speed up text entry process.

\subsection{Keyboard Layout Design Improvement}
HeadText follows an alphabetical order considering that some users 
may not be a master of QWERTY keyboard. However, for some users who are familiar with computers, it may 
be much easier for them to use QWERTY board. Therefore, in the future, we may explore 
the performance of QWERTY keyboard. The number of keys on the keyboard could also be optimized. 
The current key design is a reasonably designed keyboard layout to demonstrate the feasibility of 
HeadText. However, we see it as a long-term research process to keep improving our 
keyboard layout design. For instance, we could reduce the size of keys and increase the number of keys to 
further reduce text input ambiguity. What's more, head gesture recognition result could be improved using 
a static decoder like TipText\cite{10.1145/3332165.3347865}, and therefore text entry speed could also be improved. 

\subsection{Sensor and Prototype Form Factor}
Our prototype aims to demonstrate the feasibility of our idea. There is enough space for us to further 
improve our form factor such as decreasing the size of our prototype and using more flexible material 
to improve user experience of wearing the prototype. In addition,
HeadText prototype is based on a motion sensor -- Inertial Measurement Unit, which already exists in the 
COTS(Commercial Off-the-Shelf) device like smart earbuds. Therefore, there is potential that our technique 
could be used in existing electronic devices directly. Additionally, we could explore other sensing techniques 
like acoustic sensing for hands-free text entry in the future.

\subsection{User Study}
Although using head gestures to simulate text entry is reasonable and necessary in the social study, we admit 
that it is still better to have a real text entry study. In the future, we may use a wireless prototype to connect 
to the computer and use smart glasses to provide visual feedback or a speaker in the earbud to provide audio feedback 
to participants for auto-complete words notification and word selection. In this case, we could compare the difference of social acceptability
of real or simulated text entry.

Additionally, we have a noise gesture classifier which works well in our study. However, our text entry evaluation is in a sitting scenario.
Whether users' walking or eating may affect text entry performance continues to be unknown, since these 
noise scenarios may distract users from focusing on their current text entry and therefore nonstandard 
gestures may be recognized by mistake. Therefore, we will ask participants to evaluate text entry in a more complicated 
scenario which is more closed to the real life.

\section{Conclusion}
In this paper, we present a gesture-based hands-free technique to support users utilizing 
head gestures for text input wearing a smart earpiece using motion sensing. A 10-participant 
user study is conducted to understand users' head movements and a second study for text entry proves the 
feasibility of HeadText. Moreover, we investigate the social acceptability and privacy-protection ability of HeadText 
through a social study. We finish our paper with discussions about limitations and future work of our system. 
We believe that our technique may serve as important groundwork for future text entry on smart devices.

\bibliographystyle{ACM-Reference-Format}
\bibliography{HeadText}

\end{document}